%% file: main.tex
\documentclass[twocolumn]{aastex631}

\usepackage{newtxtext,newtxmath}

\usepackage{bm}

\usepackage{graphicx}	
\usepackage{amsmath}	
\usepackage[acronym,shortcuts]{glossaries} 
\usepackage{booktabs}
\usepackage{multirow}
\glsdisablehyper 
\usepackage{ulem} 
\usepackage{xspace}
\newcommand{\Rsun}{\ensuremath{\,\rm{R}_{\odot}}\xspace}

\newcommand{\Msun}{\ensuremath{\,\rm{M}_{\odot}}\xspace}

\newcommand{\Lsun}{\ensuremath{\,\rm{L}_{\odot}}\xspace}

\newcommand{\yr}{\ensuremath{\,\mathrm{yr}}\xspace}
\newcommand{\Msunyr}{\ensuremath{\,\mathrm{M}_\odot\,\mathrm{yr}^{-1}}\xspace}

\newcommand{\MESA}{{\scshape mesa}\xspace }
\newcommand{\COMPAS}{{\scshape compas}\xspace }

\newacronym{MS}{MS}{main-sequence}
\newacronym{ZAMS}{ZAMS}{zero-age main-sequence}
\newacronym{KH}{KH}{Kelvin-Helmholtz}
\newacronym{HR}{H-R}{Hertzsprung-Russell}
\newacronym[longplural={high-mass X-ray binaries}, plural={HMXBs}]{HMXB}{HMXB}{high-mass X-ray binary}
\newacronym[longplural={Be X-ray binaries}, plural={BeXRBs}]{BeXRB}{BeXRB}{Be X-ray binary}
\newacronym{BH}{BH}{black hole}
\newacronym{BBH}{BBH}{binary black hole}
\newacronym{RLOF}{RLOF}{Roche-lobe overflow}

\shorttitle{Expansion of accreting MS stars}
\shortauthors{Lau et al.}
\graphicspath{{./}{figs/}}

\begin{document}

\title{Expansion of accreting main-sequence stars during rapid mass transfer}

\correspondingauthor{Mike Y. M. Lau}
\email{mike.lau@h-its.org}

\author[0000-0002-6592-2036]{Mike Y. M. Lau}
\affiliation{Heidelberger Institut f\"{u}r Theoretische Studien, Schloss-Wolfsbrunnenweg 35, 69118 Heidelberg, Germany}
\affiliation{School of Physics and Astronomy, Monash University, Clayton, Victoria 3800, Australia}
\affiliation{OzGrav: The ARC Centre of Excellence for Gravitational Wave Discovery, Australia}

\author[0000-0002-8032-8174]{Ryosuke Hirai}
\affiliation{School of Physics and Astronomy, Monash University, Clayton, Victoria 3800, Australia}
\affiliation{OzGrav: The ARC Centre of Excellence for Gravitational Wave Discovery, Australia}

\author[0000-0002-6134-8946]{Ilya Mandel}
\affiliation{School of Physics and Astronomy, Monash University, Clayton, Victoria 3800, Australia}
\affiliation{OzGrav: The ARC Centre of Excellence for Gravitational Wave Discovery, Australia}

\author[0000-0002-1556-9449]{Christopher A. Tout}
\affiliation{Institute of Astronomy, University of Cambridge, Madingley Road, Cambridge, CB3 0HA, United Kingdom}
\affiliation{School of Physics and Astronomy, Monash University, Clayton, Victoria 3800, Australia}

\begin{abstract}
	Accreting main-sequence stars expand significantly when the mass accretion timescale is much shorter than their thermal timescales. This occurs during mass transfer from an evolved giant star onto a main-sequence companion in a binary system, and is an important phase in the formation of compact binaries including X-ray binaries, cataclysmic variables, and gravitational-wave sources. In this study, we compute 1D stellar models of main-sequence accretors with different initial masses and accretion rates. The calculations are used to derive semi-analytical approximations to the maximum expansion radius. We assume that mass transfer remains fully conservative as long as the inflated accretor fits within its Roche lobe, leading stars to behave like hamsters, stuffing excess material behind their expanding cheeks. We suggest a physically motivated prescription for the mass growth of such ``hamstars'', which can be used to determine mass-transfer efficiency in rapid binary population synthesis models. With this prescription, we estimate that progenitors of high-mass X-ray binaries and gravitational-wave sources may have experienced highly non-conservative mass transfer. In contrast, for low-mass accretors, the accretion timescale can exceed the thermal timescale by a larger factor without causing significant radial expansion.
\end{abstract}



\input{intro.tex}

\input{methods.tex}

\input{results.tex}
\input{discussion.tex}

\input{conclusion.tex}

\begin{acknowledgments}
  We thank the referee, Dr. Jeff Andrews, for helpful suggestions that have improved the manuscript. This work has also benefited from useful discussions with Reinhold Willcox, Philipp Podsiadlowski, and the COMPAS Team. We further thank Isobel Romero-Shaw for sharing data from her binary population synthesis study \citep{Romero-Shaw+23}. Stellar models were calculated using the OzSTAR national facility at the Swinburne University of Technology. The OzSTAR program receives funding in part from the Astronomy National Collaborative Research Infrastructure Strategy (NCRIS) allocation provided by the Australian Government, and from the Victorian Higher Education State Investment Fund (VHESIF) provided by the Victorian Government. M.Y.M.L. has been supported by an Australian Government Research Training Program (RTP) Scholarship, a Monash Postgraduate Publication Award, and a Croucher Fellowship. I.M. is a recipient of the Australian Research Council Future Fellowship FT190100574. C.A.T. thanks Churchill College for his fellowship. Parts of this research were supported by the Australian Research Council Centre of Excellence for Gravitational Wave Discovery (OzGrav), through project number CE170100004.
\end{acknowledgments}

%

\vspace{5mm}


\software{MESA r15140 \citep{Paxton+11,Paxton+13,Paxton+15,Paxton+18,Paxton+19,Jermyn+23}, COMPAS v02.41.04 \citep{Stevenson+17,Vigna-Gomez+18,TeamCOMPAS+22}. MESA inlists, subroutines, and COMPAS configuration files used in this work can be downloaded at \url{https://doi.org/10.5281/zenodo.10453751}.}


\bibliography{bibliography.bib}{}
\bibliographystyle{aasjournal}



\end{document}

%% file: intro.tex
\section{Introduction}
\label{intro}
During mass transfer in close binary stars, the accreting star may receive material at a much higher rate than it can thermally accept. For case B and case C mass transfer, the giant donor loses its envelope on its thermal timescale, which is typically many orders of magnitude shorter than the \ac{MS} accretor's thermal timescale. Past studies have found that such rapid accretion causes the stellar envelope to become overluminous and inflated, and may in many cases lead to contact \citep{Benson70,Ulrich+Burger76,Flannery+Ulrich77,Kippenhahn+Meyer-Hofmeister77,Neo+77,Packet+deGreve79}. The degree of expansion can drastically alter the outcome of \ac{RLOF}, ranging from highly-conservative mass transfer with mild expansion to coalescence, contact, or common-envelope interaction \citep{Paczynski76,Ivanova+13} with extreme inflation.


Yet, it is common for binary evolution models to neglect the accretor's response or to not model it self-consistently even when the donor star has been modelled in detail. In rapid binary population synthesis, it is common to set the accreted mass to be a fixed, arbitrary fraction of the mass lost by the donor. Another common but more elaborate approach is to restrict accretion according to the properties of the accretor at the onset of mass transfer \citep[][see Section \ref{sec:popsynth}]{Hurley+02}. However, this does not self-consistently account for the faster thermal relaxation that takes place when the accretor inflates and becomes overluminous. Accounting for these effects is possible when the donor and accretor structures are evolved during mass transfer \citep[e.g.,][]{Yungelson73,Flannery+Ulrich77,Nelson+Eggleton01,deMink+07,Deschamps+13,Fragos+23}, but such calculations are computationally expensive, and so exploring large parameter spaces still requires phenomenological, recipe-based approaches such as those used in rapid population synthesis.

We extend past studies by calculating a set of accreting \ac{MS} stellar models across a larger parameter space of initial masses, $M_0$, and constant accretion rates, $\dot{M}$. We do not account for stellar rotation, ignoring its impact on the stellar structure and the possibility of limiting accretion once the star is close to critical rotation (see Section \ref{sec:methods}). Based on these results, we provide simple prescriptions to determine the extent of accretor inflation. They may be used to estimate mass-transfer efficiency.

In Section \ref{sec:methods}, we describe our stellar models and outline their assumptions. We present calculation results in Section \ref{sec:results}, first describing the overall accretor evolution, followed by providing semi-empirical relations used to construct a formula for the maximum radius as a function of $M_0$ and $\dot{M}$. In Section \ref{sec:popsynth}, we apply these results to binary evolution models, formulating a prescription to calculate mass-transfer efficiency. We then highlight possible implications for the formation of high-mass X-ray binaries (\ref{subsec:BHHMXB}), Be X-ray binaries (\ref{subsec:BeXRBs}), and gravitational-wave sources (\ref{subsec:BBHs}). We discuss observational counterparts for binaries during rapid mass transfer in Section \ref{subsec:observations} and mention the potential effects of accretor rotation in Section \ref{subsec:rotation}. Section \ref{sec:conclusions} concludes this study.

%% file: methods.tex
\section{methods}
\label{sec:methods}
We compute spherically-symmetric, accreting stellar models using the stellar evolution code \MESA \citep[r15140,][]{Paxton+11,Paxton+13,Paxton+15,Paxton+18,Paxton+19,Jermyn+23}. These models are hydrostatic and do not account for rotation and mass loss. To generate a starting \ac{ZAMS} model, we use \MESA's default options to initialise a pre-\ac{MS} star of a given mass \citep{Paxton+11} with solar metallicity $Z=0.0142$ \citep{Asplund+09}. Evolution is stopped when the central hydrogen mass fraction begins to decrease. We then linearly increase the accretion rate over 20 time-steps to the desired value. Actual mass transfer rates in binaries are very non-uniform \citep{Flannery+Ulrich77} and depend on uncertain assumptions such as the angular momentum carried away by lost material and the donor star's structural response to mass loss \citep{Hjellming+Webbink87,Soberman+97}. Despite this, we choose to adopt constant accretion rates in order to study the accretor response in isolation \citep[e.g.,][]{Kippenhahn+Meyer-Hofmeister77}.

We use the \cite{Ledoux47} criterion to determine convective stability and adopt the \cite{Cox+Giuli68} description of mixing-length theory \citep{Bohm-Vitense58} with a mixing-length parameter of $\alpha_\mathrm{MLT} = 2$. We allow step overshooting above the hydrogen-burning core with the parameters\footnote{The inlists and subroutines used for generating our models are available at \url{https://doi.org/10.5281/zenodo.10453751}.} $f_\mathrm{ov}=0.33$ and $f_\mathrm{ov,0}=0.05$ \citep{Brott+11}. No mass loss is included in the models. We adopt an atmospheric pressure that is larger than the default by adjusting \MESA's \texttt{Pextra\_factor} parameter from 1 (corresponding to an Eddington grey atmosphere) to 1.36. This aids convergence for the most rapidly accreting and initially more massive models.

The added material is given the same composition and specific entropy as the surface. The latter assumption was made in all previous studies of accretor expansion during rapid mass transfer, implying a process that removes the gravitational potential energy of added mass that is in excess to the stellar surface. One possibility is that during \ac{RLOF}, the accretion stream has sufficient angular momentum to form an accretion disk around the accretor, allowing this excess energy to be dissipated during inward migration. In the case of direct-impact accretion, efficient shock cooling may take via a hot spot formed by the accretion stream \citep{Ulrich+Burger76}.

Our calculations assume spherically-symmetric accretion. In reality, a \ac{MS} star can spin up to break-up rotational speeds by accreting a few percent of its mass with the specific angular momentum of its surface \citep{Packet81}. Whether the star can continue accreting at this stage and regulate its nearly critical rotation is a major uncertainty \citep{Petrovic+05,deMink+13,Deschamps+13}. Models of accretion onto rapidly rotating stars find that angular momentum can be removed through an accretion disk and star-disk boundary layer to enable sustained accretion near critical rotation \citep{Paczynski91,Popham+Narayan91,Colpi+91,Bisnovatyi-Kogan93}. Furthermore, tides raised on the accretor may regulate rotation by transferring angular momentum from its inflated envelope or outer edge of an accretion disk to the orbit, potentially averting breakup. Magnetic disk-locking is another mechanism for coupling the accretor with a surrounding accretion disk \citep{Armitage+Clarke96,Stepien00,Dervisoglu+10}. We do not account for any of these effects. Instead, we study the impact of stellar inflation in isolation. 

Finally, the efficiency of convection in superadiabatic layers of near-Eddington envelopes is highly uncertain \citep{Sanyal+15}. The usual assumptions of mixing-length theory are no longer valid and effects such as wave-driven transport, porosity, and super-Eddington winds may completely modify the sizes and luminosities of these inflated envelopes \citep{Shaviv01a,Shaviv01b,Owocki+04}. In light of these uncertainties, this Letter represents only a first step towards understanding the qualitative effect of accretor inflation on mass transfer. The framework we provide can be refined with more sophisticated stellar models in the future.

\subsection{Simulation grid}
\label{subsec:grid}
\begin{figure}
    \includegraphics[width=\linewidth]{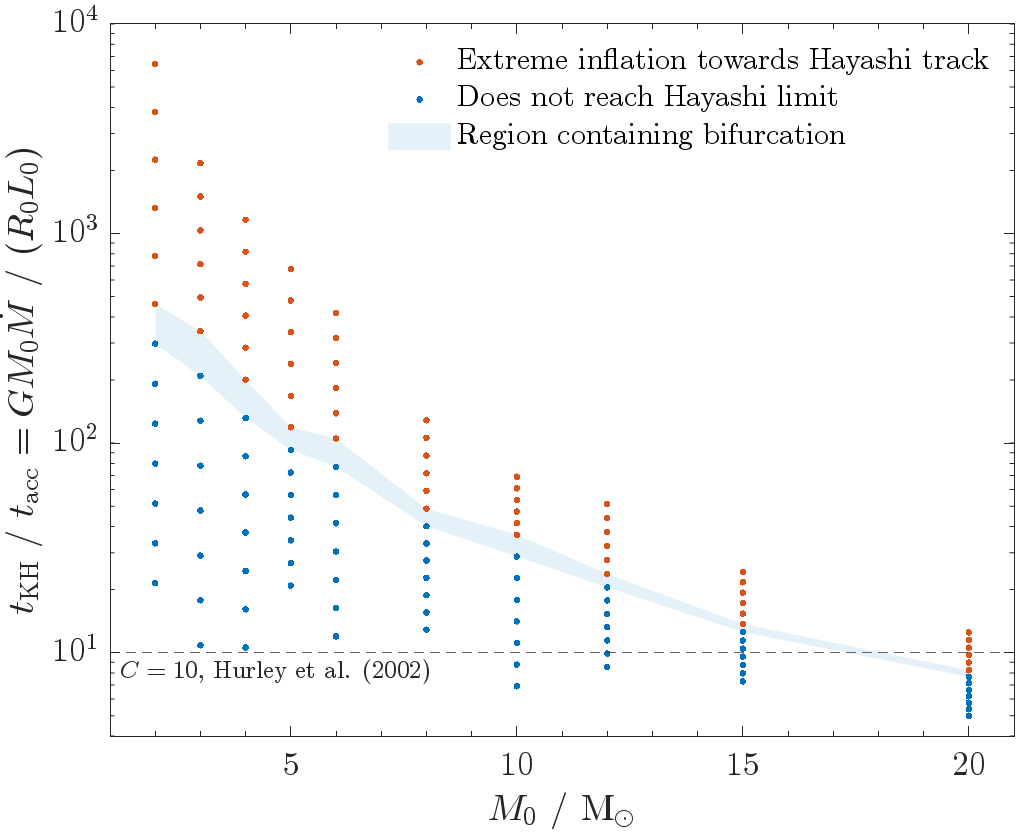}
        \caption{The parameter space explored in this study. Each point represents one of 130 models we computed. The abscissa shows the initial mass while the ordinate parametrises $\dot{M}$ in terms of the ratio of the accretor's initial Kelvin-Helmholtz timescale to the accretion timescale. Red markers are accretors that experience extreme inflation towards the Hayashi line and the blue markers are accretors that do not. The bifurcating timescale ratio is contained within the shaded region. That this region is mainly located where $t_\mathrm{KH}/t_\mathrm{acc} \gg 10$ suggests a large range of $\dot{M}$ could result in more conservative mass transfer than assumed by \cite{Hurley+02} (see Section \ref{sec:popsynth}).}  
    \label{fig:Cfactor}
\end{figure}

We compute a total of 130 accreting models consisting of ten different initial masses ($M_0/ \Msun\in \{2,3,4,5,6,8,10,12,15,20\}$) each with 13 different logarithmically-spaced accretion rates, $\dot{M}$. The accretion rates we have chosen mean that the accretor's Kelvin-Helmholtz timescale is a factor of a few to thousands of times longer than the accretion timescale, as shown in Figure \ref{fig:Cfactor}. Our selection of initial masses focuses on more massive models because the high accretion rates we explore require massive donors. This usually leads to dynamically unstable mass transfer that results in a common envelope or merger for low-mass accretors. For each $M_0$, the smallest $\dot{M}$ was selected so as to cause mild inflation by a factor of a few. The highest $\dot{M}$ was chosen to limit the total accreted mass to about $100\Msun$ upon reaching maximum radius, which can be as large as several thousand solar radii. As we will show in Section \ref{sec:results}, the six highest values of $\dot{M}$ explore a regime where the accretor promptly reaches a maximum luminosity and expands along a Hayashi line.

%% file: results.tex
\section{Results}
\label{sec:results}
\begin{figure*}
    \includegraphics[width=\linewidth]{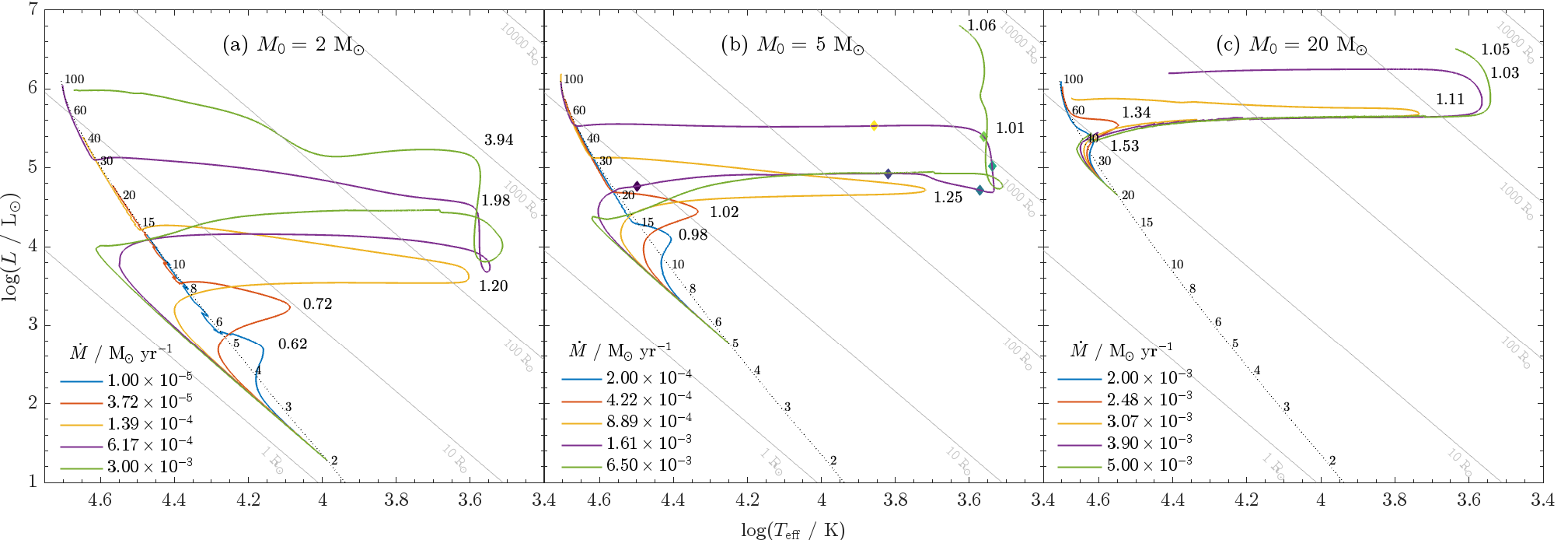}
    \caption{H-R diagrams showing the evolution of rapidly accreting main-sequence stellar models with initial masses $M_0=2,5,20\Msun$ (panels (a)-(c)). Each panel shows five different accretion rates as indicated by the legend. The timescale ratio $t_\mathrm{th}/t_\mathrm{acc}$ (defined by equation \ref{eq:ratio}) is shown for each model at maximum radius, demonstrating that contraction begins when it decreases to approximately unity. The dotted line in each panel shows the zero-age main sequence with masses labelled in units of \Msun. In panel (b), stellar profiles at the diamond markers along the purple track are shown in Figure \ref{fig:prof}.}
    \label{fig:HR}
\end{figure*}

Figure \ref{fig:HR} shows the evolution of the accretor in a \ac{HR} diagram for initial masses $M_0\in\{2,5,20\}\Msun$ and for an illustrative subset of accretion rates. The general behaviour is consistent with previous calculations \citep{Ulrich+Burger76,Neo+77,Kippenhahn+Meyer-Hofmeister77}. Namely, in the limit of low accretion rates that allow the star to maintain thermal equilibrium, the accretor ascends the \ac{ZAMS} (dotted line) while assuming a position in accordance with its current mass. While for larger accretion rates, the star inflates and veers towards the red before contracting and ascending the \ac{ZAMS}. Upon accretion, the surface luminosity is driven up almost immediately and adiabatically towards the accretion luminosity, $L_\mathrm{acc} = GM\dot{M}/R$. This occurs at a fixed radius, and so the effective temperature increases.

\begin{figure}
    \includegraphics[width=\linewidth]{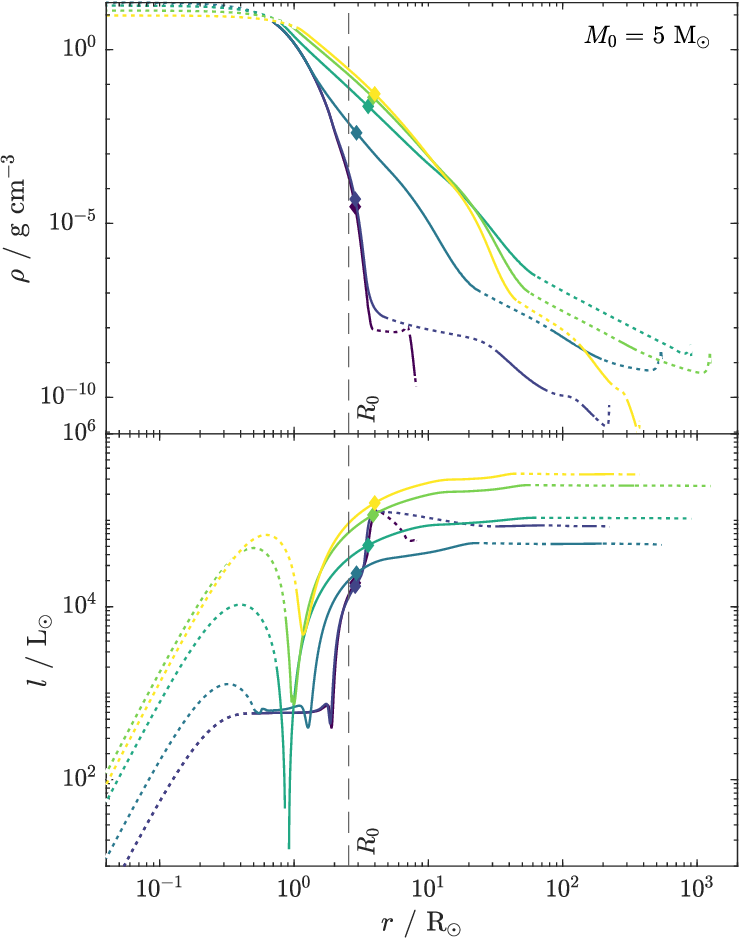}
    \caption{Density (top panel) and luminosity (bottom panel) profiles of a $M_0=5\Msun$ model accreting at a rate of $\dot{M}=1.61\times10^{-3}\Msunyr$ at different stages of its evolution. The line colour changes from indigo to yellow from earlier to later models. Dashed lines indicate convective regions. The position of the models on the H-R diagram are marked on the purple line in Figure \ref{fig:HR}(b). The vertical dashed line shows the initial stellar radius, $R_0=2.6\Rsun$, and the diamond markers show $R_\mathrm{eff}$ as defined by equation (\ref{eq:Reff}).}
    \label{fig:prof}
\end{figure}

Figure \ref{fig:prof} shows density and luminosity profiles of a $M_0=5\Msun$ accretor accepting mass at a rate of $\dot{M}=1.61\times10^{-3}\Msunyr$. We include profiles at different stages of its evolution, indicated by markings along the purple track in Figure \ref{fig:HR}(b). The upper panel illustrates the general finding that the accreted material accumulates in a low-density envelope outside the initial stellar radius. This is responsible for rightward evolution in the \ac{HR} diagram, with the extent of inflation increasing with $\dot{M}$. The lower panel shows concentrated release of gravitational potential energy near the base of the envelope at a dip in the luminosity profile ($l(r)$). Simultaneously, the deep inner layers expand and absorb energy ($\partial l / \partial r < 0$) as they adjust to the growing accretor mass.

Beyond a threshold $\dot{M}$, the stellar model experiences a maximum luminosity that depends on $M_0$ and expands towards a minimum temperature, $\log(T_\mathrm{min}$~/~K$)\approx 3.6$, set by the Hayashi limit\footnote{Throughout this paper, $\log$ denotes logarithm with base 10.}. This is illustrated by the green and purple lines in Figures \ref{fig:HR}(a) to (c). This expansion is rapid and takes place before a large amount of mass has been added. Subsequently, the accretor expands with roughly constant effective temperature, ascending the Hayashi line as a red giant with a convective envelope.

\subsection{Maximum accretor radius}
As the accretor becomes more luminous, it more quickly transports and radiates away the excess gravitational energy of the added material. The star eventually reaches a maximum radius and contracts. We are able to express the approximate condition for contraction as a balance between the accretion timescale, $t_\mathrm{acc} = M/\dot{M}$, and a thermal-adjustment timescale, $t_\mathrm{th}$:
\begin{align}
    \frac{t_\mathrm{th}}{t_\mathrm{acc}} = \frac{GM\dot{M}}{R_\mathrm{eff}(M)L} = 1,
    \label{eq:ratio}
\end{align}
where $M$ and $L$ are the instantaneous stellar mass and luminosity, and
\begin{align}
    \frac{R_\mathrm{eff}(M)}{\Rsun} = 2\bigg(\frac{M}{\Msun}\bigg)^{0.22}.
    \label{eq:Reff}
\end{align}
The expression for $t_\mathrm{th}$ differs from the Kelvin-Helmholtz timescale in the replacement of the stellar radius, $R$, by an effective radius $R_\mathrm{eff}$ that represents the radial scale at which excess gravitational potential energy is released relative to the equilibrium structure. This radius weakly increases with stellar mass and is much smaller than the inflated envelope. Thus, $t_\mathrm{th}/t_\mathrm{acc}\approx 1$ is equivalent to the luminosity being able to carry away energy at the same rate, $L_\mathrm{settling}$, at which it is released through gravitational settling to the radius $R_\mathrm{eff}$. Similar criteria have been determined for rapid accretion onto protostars in models of supermassive stellar formation \citep{Sakurai+15,Nandal+23}.

In Figure \ref{fig:HR}, we label $t_\mathrm{th}/t_\mathrm{acc}$ for each evolution sequence upon reaching maximum radius, demonstrating that it decreases to approximately unity for the plotted models. After this point, the star contracts towards a thermally-relaxed structure, reaching the \ac{ZAMS} at the position appropriate to its current mass. Because of its much larger mass, which implies a larger luminosity, the star remains in thermal equilibrium despite continued accretion.


\subsection{Mass-luminosity relation of inflated envelopes}
\label{subsec:ML_relation}
\begin{figure}
    \includegraphics[width=\linewidth]{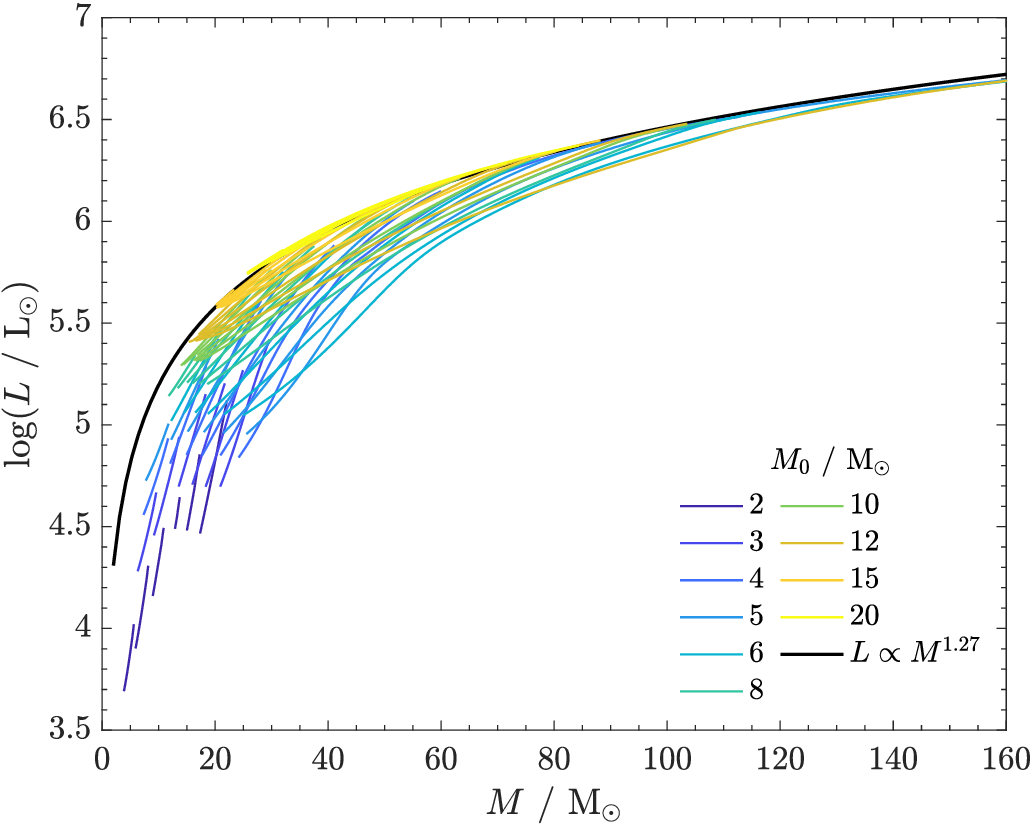}
    \caption{Mass-luminosity relation for rapidly accreting models during their expansion along the Hayashi line up to their maximum radii. Lines of the same colour show models with same initial masses but different accretion rates. The black line shows a power-law fit to models with larger initial masses.}
    \label{fig:L_vs_M}
\end{figure}

Some models that reach a maximum luminosity and expand along the Hayashi line are found to satisfy an approximate mass-luminosity relation that is independent of the initial mass and accretion rate, similar to the Eddington luminosity. Figure \ref{fig:L_vs_M} demonstrates this by plotting $\log(L$~/~L$_\odot)$ against $M$ for these models, but restricted to the constant-temperature expansion phase on the Hayashi line. We obtain a fit of $\log(L$~/~L$_\odot) = 1.27\log(M$~/~M$_\odot) + 3.93$ to the more luminous and massive models. Similar mass-luminosity relations have been found for near Eddington luminosity stellar envelopes in other contexts, such as Wolf-Rayet stars \citep{Langer89,Schaerer+Maeder92}, rapidly accreting protostars \citep[e.g.,][]{Stahler+86,Hosokawa+Omukai09}, and supernova companion stars that have been inflated by ejecta impact \citep{Ogata+21,Hirai23}.

The mass-luminosity relation breaks down for lower $M_0$ and/or higher $\dot{M}$, particularly toward lower luminosities. In these cases, the luminosity is influenced by $\dot{M}$. We attribute this to the fact that the inflated envelope has not yet reached a steady state. Therefore, our power-law fit does not provide a reliable estimate of the turnaround luminosity for models with $M_0 \lesssim 4\Msun$. However, we emphasise again that such low-mass systems are not the focus of our study, because the high accretion rates under consideration require massive donors, usually resulting in the loss of dynamical stability leading to a common-envelope inspiral or merger.

\begin{figure}
    \includegraphics[width=\linewidth]{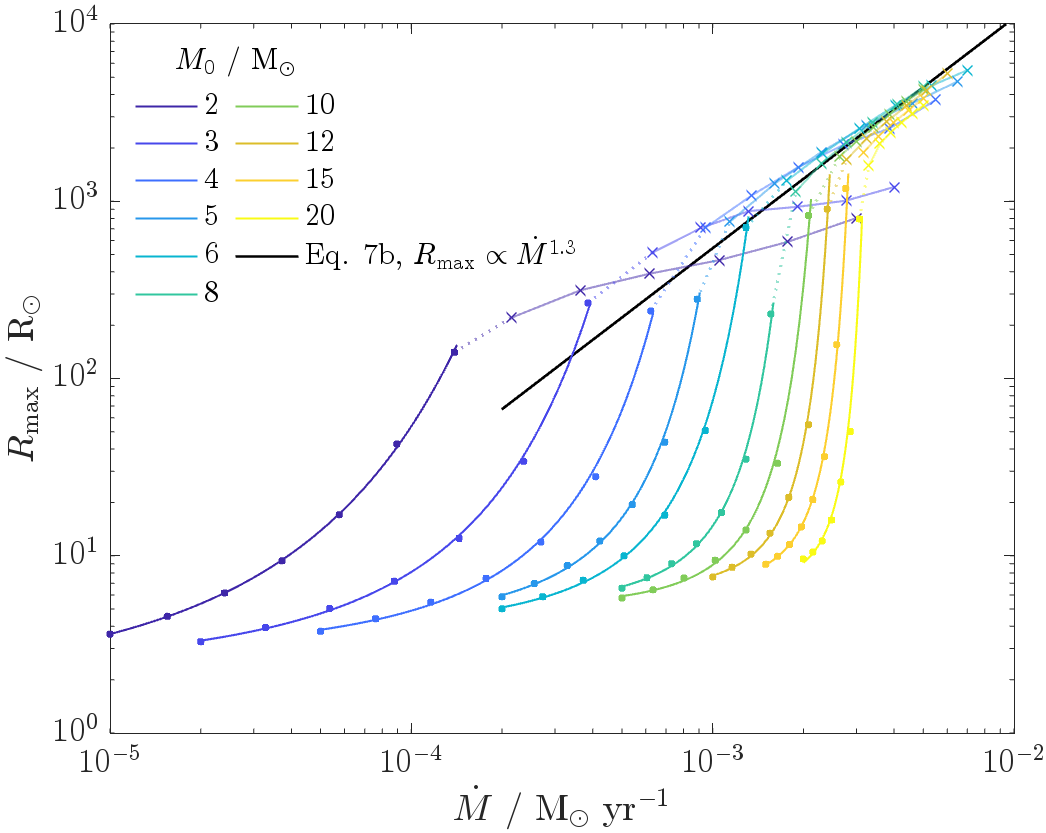}
    \caption{The maximum radius, $R_\mathrm{max}$, as a function of accretion rate, $\dot{M}$, reached by accretors with different initial masses, $M_0$. Models with lower $\dot{M}$ that do not expand to their Hayashi limits are plotted as dot markers and lie close to the empirical fits we provide (equation \ref{eq:rmax_subhay}). High-$\dot{M}$ models that expand along the Hayashi line (cross markers) do not retain memory of their initial masses, and lie close to our semi-analytical formula (equation \ref{eq:Rmax_scaling}, black line), although there are deviations for $M_0 < 4\Msun$.}
    \label{fig:Rmax_vs_mdot}
\end{figure}

With a mass--luminosity relation and a criterion for maximum expansion (equation (\ref{eq:ratio})), one can solve for the maximum radius reached by a stellar model that ascends the Hayashi line as the solution to the set of equations,
\begin{align}
    L &= \frac{GM^{1-\zeta} \mathrm{M}_\odot^\zeta \dot{M}}{\lambda\Rsun}, \label{eq:a} \\
    L/\Lsun &= k_1(M/\Msun)^{k_2}, \label{eq:b} \\
    L &= 4\pi \sigma R^2 T_\mathrm{eff}^4, \\
    T_\mathrm{eff} &= T_\mathrm{min}, \label{eq:d}
\end{align}
where $\sigma$ is the Stefan-Boltzmann constant, $k_1 = 10^{3.93}$ and $k_2 = 1.27$ are fitting coefficients to the mass-luminosity relation, $\log(T_\mathrm{min}$ / K$) = 3.6$ is the approximately constant temperature along the Hayashi line, $\zeta = 0.22$ is the power-law scaling of the effective settling radius with mass, and $\lambda = 2$ is a multiplicative factor to the settling radius. The solution to equations (\ref{eq:a}) to (\ref{eq:d}) is
\begin{subequations}
\begin{align}
    \frac{R_\mathrm{max}}{\Rsun} &=
        \Bigg[ \bigg(\frac{L_1}{k_1\Lsun}\bigg)^{1-\zeta}
        \bigg(\frac{L_2}{L_1}\bigg)^{k_2} \Bigg]^{\frac{1}{2[k_2-(1-\zeta)]}}
        \label{eq:rmax} \\
        &= 540~\bigg(\frac{\dot{M}}{10^{-3}\Msunyr}\bigg)^{1.3},
        \label{eq:Rmax_scaling}
\end{align}
\end{subequations}
where
\begin{align}
    L_1 &= 4\pi\sigma\Rsun^2T_\mathrm{min}^4~~~\mathrm{and}\\
    L_2 &= G\mathrm{M}_\odot \dot{M}/(\lambda\Rsun).
\end{align}
Figure \ref{fig:Rmax_vs_mdot} compares equation (\ref{eq:Rmax_scaling}, black line) to the actual $R_\mathrm{max}$ recorded in our calculations. It shows a good match for rapidly accreting models that ascend the Hayashi line (cross markers). That these models lie approximately on a single line reflects the fact that the turnaround luminosity carries no memory of the initial mass. There are noticeable deviations for the $M_0 = 2$ and $3\Msun$ models, because the mass-luminosity relation applies poorly to these initial masses. 

For models with lower accretion rates (dot markers), which do not reach the Hayashi limit, we have to fit $R_\mathrm{max}$ to both $M_0$ and $\dot{M}$. We choose the functional form
\begin{align}
    \log\bigg[ \log \biggl( \frac{R_\mathrm{max}}{R_0} \biggr) \bigg] = 10^a \biggl(\frac{\dot{M}}{\Msunyr} \biggr)^b - c,
    \label{eq:rmax_subhay}
\end{align}
where $R_0$ is the \ac{ZAMS} radius and the fitting coefficients $a$, $b$, and $c$ are displayed in Table \ref{tab:fits} for each choice of $M_0$. These fits are plotted as the coloured, thick solid lines in Figure \ref{fig:Rmax_vs_mdot}. For comparison, \cite{Pols+Marinus94} also used a double logarithm, but fitted to a linear function of $\log\dot{M}$ instead of a power law. Their fit was based on a limited number of models computed by \cite{Kippenhahn+Meyer-Hofmeister77}, \cite{Neo+77}, and \cite{Packet+deGreve79}.

Figure \ref{fig:Rmax_vs_mdot} also shows that for more massive accretors, $R_\mathrm{max}$ is highly sensitive to $\dot{M}$. For the $M_0=20\Msun$ model, $R_\mathrm{max}$ increases from 9\Rsun at $\dot{M}=2\times 10^{-3}\Msunyr$ to complete expansion towards the Hayashi limit (2000\Rsun) after increasing $\dot{M}$ by just a factor of 1.5. This could be because the unperturbed luminosities of more massive \ac{MS} accretors are closer to their critical luminosities. Similarly, \cite{Neo+77} state that more massive accretors need to be closer to their Eddington-limited accretion rates to significantly inflate. The response of a more massive star towards rapid mass inflow therefore appears to bifurcate at a critical inflow rate that controls whether mild expansion or extreme inflation along the giant branch takes place. This bifurcation is located within the shaded band in Figure \ref{fig:Cfactor}, which divides models that expand all the way towards the Hayashi limit (red markers) from those that do not (blue markers).


\begin{table}
    \caption{$R_\mathrm{max}$ fitting coefficients}
    \label{tab:fits}
    {
        \centering
        \begin{tabular}{@{}llll@{}}
            \toprule
            $M_0$ / \Msun & $a$    & $b$    & $c$  \\ \midrule
            2             & 0.920  & 0.146  & 1.99 \\
            3             & 1.17   & 0.276  & 1.36 \\
            4             & 1.52   & 0.426  & 1.12 \\
            5             & 2.58   & 0.844  & 0.722 \\
            6             & 2.22   & 0.731  & 0.924 \\
            8             & 3.77   & 1.37   & 0.727 \\
            10            & 3.73   & 1.36   & 0.894 \\
            12            & 5.45   & 2.06   & 0.763 \\
            15            & 5.97   & 2.34   & 0.856 \\
            20            & 6.26   & 2.42   & 1.19 \\
            \bottomrule
        \end{tabular}
        \par
    }
    \tablecomments{Fitting coefficients for the maximum radius, $R_\mathrm{max}(\dot{M})$, computed using equation (\ref{eq:rmax_subhay}) for an accretor that does not reach the Hayashi limit.}
\end{table}

%% file: discussion.tex
\section{Discussion}
\label{sec:discussion}

\subsection{Prescription for mass-transfer efficiency in binary modelling}
\label{sec:popsynth}
Our results have important implications for modelling binary stellar evolution. The extent of radius inflation during mass transfer influences the fraction of mass lost by the donor that is added to the accretor. This is called the mass-transfer or accretion efficiency, commonly denoted $\beta$. It determines the mass growth of the accretor and the resulting orbital separation. However, its value is uncertain and is commonly set to an arbitrary constant in binary population synthesis calculations. Another common approach involves setting $\beta$ to be the ratio $\dot{M}_\mathrm{a,max} / \dot{M}_\mathrm{d}$ of the accretor's thermally-limited maximum mass acceptance rate to the donor's mass-loss rate \citep[e.g.,][]{PortegiesZwart+Verbunt96,Hurley+02,Belczynski+08,Giacobbo+18,TeamCOMPAS+22},
\begin{align}
    \beta = \min\biggl(1,\frac{\dot{M}_\mathrm{a,max}}{\dot{M}_\mathrm{d}}\biggr)
    = \min\bigg(1, C \frac{R_\mathrm{a} L_\mathrm{a}}{GM_\mathrm{a}\dot{M}_\mathrm{d}}\bigg),
    \label{eq:beta}
\end{align}
where $M_\mathrm{a}$, $R_\mathrm{a}$, and $L_\mathrm{a}$ are the accretor mass, radius, and luminosity at the onset of \ac{RLOF}, $\dot{M}_\mathrm{d}$ is the donor's mass-loss rate, and $C$ is a dimensionless parameter that accounts for increased accretion efficiency as the accretor inflates and becomes over-luminous during mass transfer. The ratio $\dot{M}_\mathrm{a,max}/\dot{M}_\mathrm{d}$ is also equal to the ratio of the accretion timescale to the initial Kelvin-Helmholtz timescale, $t_\mathrm{acc}/t_\mathrm{KH}$. Therefore, $C$ is the factor by which $t_\mathrm{KH}$ can exceed $t_\mathrm{acc}$ before mass transfer is no longer modelled as fully conservative. Many studies adopt $C=10$ following \cite{Hurley+02}.

Our results demonstrate that it is possible for accretors to expand mildly despite being fed much faster than the thermally-limited rate. For example, the 2\Msun model expands by a factor of 2.3 when fed at 20 times its thermally-limited mass acceptance rate ($\dot{M} = 10^{-5}\Msunyr$). It is then reasonable to posit that all the donated mass can be added as long as the inflated envelope remains within the accretor's Roche lobe, much like how a ravenous hamster can accept food much faster than it can ingest by stuffing most of it behind its expanding cheeks. \cite{Pols+Marinus94} adopted such a model in their population synthesis study, although their determination of the maximum accretor radius was based on a limited number of accreting stellar models published at the time. Based on this assumption, the parameter $C$ in equation (\ref{eq:beta}) can be much larger than unity, and should not be held constant because the degree of expansion depends on $M_0$ and $\dot{M}$. 

The approximate magnitude of $C$ may be indicated by the shaded region in Figure \ref{fig:Cfactor}. This contains the critical ratio $t_\mathrm{KH}/t_\mathrm{acc}$ controlling whether mild expansion or substantial inflation towards the Hayashi limit takes place. Indeed, we find that up to $M_0 = 15\Msun$, the shaded region encompasses values much greater than the nominal $C=1$ or even the commonly adopted $C=10$. This leads to a larger parameter space for fully conservative mass transfer, although the exact boundary depends on the pre-\ac{RLOF} orbital separation. On the other hand, the displayed trend suggests $C<10$ for $M_0\gtrsim20\Msun$, meaning such massive accretors may accrete even less conservatively than typically assumed. Our prescription therefore predicts both more conservative and less conservative mass transfer depending on the initial accretor mass. Although our calculations are limited to \ac{ZAMS} accretors, as a first approximation, one can assume that more evolved accretors have the same critical $t_\mathrm{KH}/t_\mathrm{acc}$ ratio. But, because of their larger radii and luminosities, this implies a larger critical $\dot{M}$ that scales as $(R/R_\mathrm{ZAMS})(L/L_\mathrm{ZAMS})$. 

\begin{figure}
    \includegraphics[width=\linewidth]{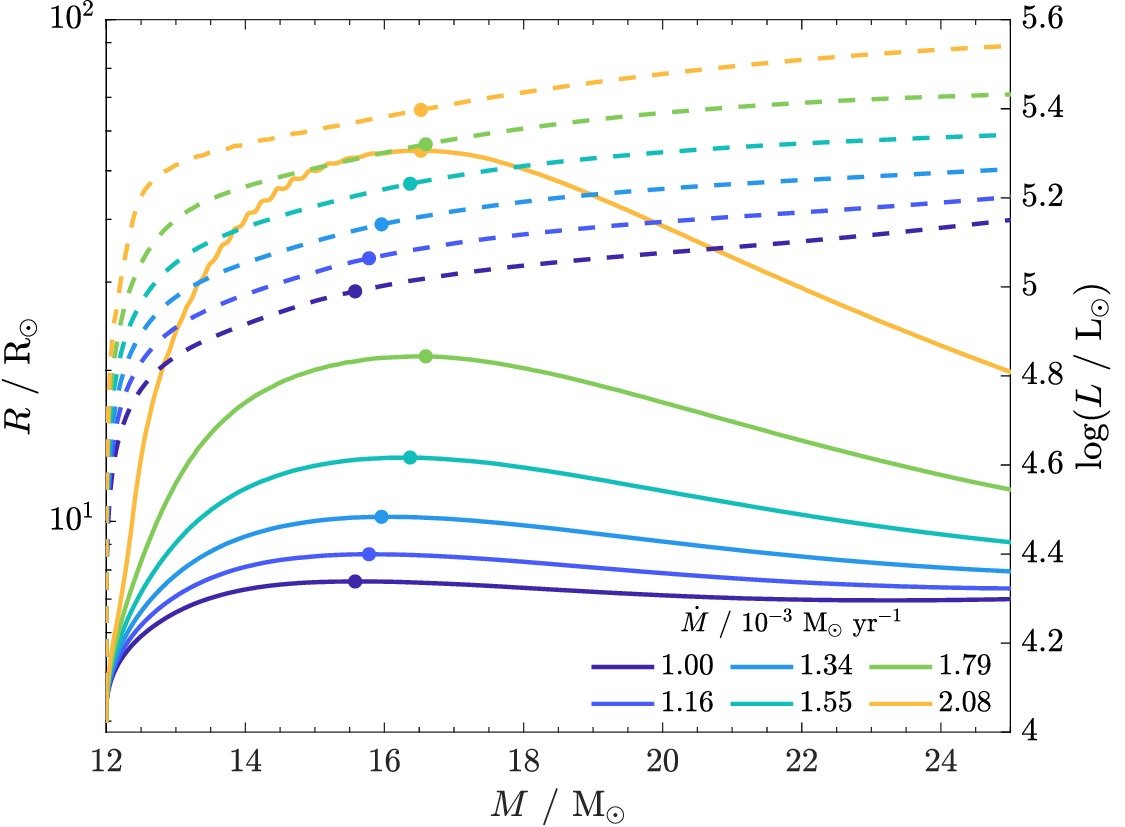}
    \caption{Radius (solid lines) and luminosity (dashed lines) evolution of an initially 12\Msun accretor. The abscissa shows the accretor mass, which is proportional to time (fully conservative accretion is assumed). Line colours correspond to different accretion rates, while the dot along the curve indicates the maximum radius point.}
    \label{fig:R_L_evolution}
\end{figure}

We now present how the total accreted mass may be determined using our fits to $R_\mathrm{max}$.
\begin{enumerate}
    \item From an averaged mass donation rate $\dot{M}_d$ and the accretor mass $M_0$ at the onset of \ac{RLOF}, calculate $R_\mathrm{max}$ with the prescriptions provided in Section \ref{sec:results}.
    \item During mass transfer, the accretor Roche radius, $R_L$, varies in response to changing orbital separation and mass ratio. Solve the differential equation governing orbital separation, first assuming $\beta=1$. Check if this results in a self-consistent solution satisfying $R_\mathrm{max} \leq R_L$ at every numerical step. If so, accept this fully conservative solution. Otherwise, proceed to step 3.
    \item Determine the maximum amount of mass, $\Delta M$, that can be added to the accretor before it fills its Roche lobe. This can be done by solving the equation $R(M_0+\Delta M) = R_L(M_0+\Delta M)$ with a suitable root finder. \citeauthor{Eggleton83}'s (\citeyear{Eggleton83}) formula, for example, can be used for $R_L$, whereas $R(M_0+\Delta M)$ may be calculated by solving equations (\ref{eq:b}) to (\ref{eq:d}). If $\Delta M$ is greater than the total mass lost by the donor, not enough mass is added to the accretor for it to actually reach $R_L$, and we have a fully conservative solution. Otherwise, proceed to step 4.
    \item Assume the mass, $\Delta M$, is added conservatively, as it is contained within the Roche lobe and eventually gravitationally settles. Also suppose that the remaining mass can be added to the Roche-filling accretor at a rate consistent with its luminosity, $L(M_0+\Delta M)$. That is, the accretion efficiency after filling the Roche lobe is $\beta = L/L_\mathrm{settling} = R_\mathrm{eff}L/(GM\dot{M}_d)$, where $L_\mathrm{settling}$ is the rate of energy release at the effective radius $R_\mathrm{eff}(M)$ at which mass gravitationally settles onto the accretor, calculated with equation (\ref{eq:Reff})\footnote{While equation (\ref{eq:Reff}) is calibrated to predict the maximum radii of accreting models, we assume, as a first approximation, that it also represents the effective settling radius for models that are still expanding.}.
\end{enumerate}
We note several potential complications that may arise from the final step. To begin with, $R(M)$ and $L(M)$ may only be determined during expansion along the Hayashi line, where equations (\ref{eq:b}) to (\ref{eq:d}) apply. If the accretor fills its Roche lobe at a hotter phase, one may instead approximate the luminosity during contact as the initial accretion luminosity, $L \approx GM_0\dot{M}_\mathrm{d}/R_0$, and evaluate the effective radius with the initial mass, $R_\mathrm{eff} \approx R_\mathrm{eff}(M_0)$. Figure \ref{fig:R_L_evolution} shows that these are reasonable approximations, with the radius and luminosity evolution of a $M_0=12\Msun$ model as an example. Firstly, upon adding mass, the luminosity rises steeply due to accretion before the star expands appreciably. This can also be seen in the \ac{HR} diagram (Figure \ref{fig:HR}). Secondly, most of the expansion takes place before a substantial amount of mass has been accrued. At most, the stellar mass increases by few tenths at maximum radius. Thus, to obtain a lower limit for mass-transfer efficiency, one can neglect this small amount of mass that in principle can be added conservatively prior to filling the Roche lobe.

Another complication in step 4 is that the accretor's Roche radius may surpass $R_\mathrm{max}$ again, whether due to the increased accretor mass or mass-ratio reversal. In such a scenario, the remaining available mass may once again be added with perfect efficiency.

\subsection{Example applications}
\label{subsec:examples}
In this section, we assess the potential impact of our mass-transfer efficiency prescription on modelling the formation of several classes of compact binaries, including \ac{BH} \acp{HMXB}, \acp{BeXRB}, and gravitational-wave sources. We do not perform a full calculation of the accreted mass as outlined in the previous section, as this is beyond the scope of this work. Instead, we estimate the extent of accretor inflation based on conditions at the initiation of mass transfer, as determined by rapid binary population synthesis models.

\subsubsection{Black-hole high-mass X-ray binaries}
\label{subsec:BHHMXB}
Recently, \citet[][hereinafter, R23]{Romero-Shaw+23} found that thermally-limited accretion (equation \ref{eq:beta}) and other common assumptions in binary modelling leads to overestimating the number of \ac{BH}-\acp{HMXB}. They proposed that formation through case B mass transfer may be suppressed if it is assumed to be more conservative. This aids the widening of the orbit upon mass-ratio reversal. A wider orbit would in turn prevent an accretion disk from forming around the \ac{BH} later in the evolutionary sequence, a process that requires the wind-feeding optical companion to have a Roche-lobe filling fraction above 80\% \citep{Hirai+Mandel21}. Observational indications of near-maximal spins in \ac{BH}-\acp{HMXB} also disfavour formation through case B mass transfer \citep{Qin+19,Neijssel+21}. 

We examine whether our proposed prescription results in more conservative case B mass transfer. Using the population synthesis output data from R23 (I. Romero-Shaw, private communication), we determined the typical conditions at the onset of case B mass transfer leading to \ac{BH}-\ac{HMXB} formation. Typical primary and secondary masses are $M_1=40\Msun$ and $M_2=10\Msun$, with an orbital separation of 110 to 180\Rsun. Assuming the primary loses a 25\Msun envelope on the Hertzsprung gap over its Kelvin-Helmholtz timescale, which varies from almost 2000\yr initially to 20\yr at maximum extent, the average mass transfer rate ranges from $10^{-2}$ to 1\Msunyr.

Our stellar models show that these mass transfer rates cause extreme inflation of unevolved accretors, extending beyond the Roche lobe. For example, a 10\Msun model accreting at $3\times10^{-3}\Msunyr$ expands to 1200\Rsun after adding just 4\Msun. Consequently, we anticipate highly non-conservative mass transfer, contrary to R23's suggestion. The binary may evolve into contact, losing mass from the $L_2$ Lagrange point instead of the commonly assumed centre of mass of the accretor. This could take away a larger specific angular momentum and increases the likelihood of entering dynamically unstable mass transfer \citep[e.g.,][]{Willcox+23}. The binary can then merge or enter a common envelope, producing a drastically different outcome than stable, thermal-timescale mass transfer. This scenario could make it less likely to form a \ac{BH}-\ac{HMXB} through case B mass transfer, and thus still explaining the observations which compelled R23 to propose conservative mass transfer in this regime. 

\subsubsection{Be X-ray binaries}
\label{subsec:BeXRBs}
Population synthesis models that assume thermally-limited accretion also appear to be in tension with properties of observed \acp{BeXRB}, which comprise a large fraction of known \acp{HMXB}. These systems consist of a neutron star and a rapidly rotating B-type star, characterised by Balmer emission lines from a surrounding decretion disk. The models tend to under-predict observed orbital periods and stellar-component masses \citep{Vinciguerra+20,Liu+24}. Similarly, it has been suggested that more conservative mass transfer is necessary for the stellar accretor to accumulate more mass and for the binary orbit to further contract.

To determine whether this may follow from our mass transfer prescription, we replicate the population synthesis calculation by \cite{Vinciguerra+20} to characterise the typical conditions of the mass transfer episode experienced by \ac{BeXRB} progenitors. We used the same population synthesis code, \COMPAS \citep[v02.41.04,][]{Stevenson+17,Vigna-Gomez+18,TeamCOMPAS+22}, and same assumptions as in their default model\footnote{The \COMPAS configuration files used in this work can be downloaded at \url{https://doi.org/10.5281/zenodo.10453751}.}. We adopt the same fixed metallicity of $Z = 0.0035$, consistent with the Small Magellanic Cloud.

We find that the donor is typically around 10\Msun and has an approximately 8\Msun envelope when it exceeds its Roche lobe during Hertzsprung gap evolution. If its envelope is removed within its Kelvin-Helmholtz timescale of about 3000\yr, the average mass transfer rate amounts to $3\times10^{-3}\Msunyr$. For a typical accretor of 3 to 10\Msun, this accretion rate would cause expansion to the Hayashi limit, exceeding $1000\Rsun$ (see Figure \ref{fig:Rmax_vs_mdot}). This would expand the accretor beyond its Roche lobe. However, with a moderately lower mass transfer rate of $10^{-3}\Msunyr$, a 5\Msun accretor would only inflate to 100\Rsun, allowing for fully conservative mass transfer within our framework. Due to this sensitivity to $\dot{M}$, we anticipate both highly conservative and highly non-conservative mass transfer to occur across a wider parameter region. This makes it possible to account for individual observed systems that are known to have accreted efficiently, such as a recently discovered Be-binary with a partially stripped primary, a possible \ac{BeXRB} progenitor \citep{Ramachandran+23}. However, a self-consistent calculation is necessary to precisely determine the impact on the observed \ac{BeXRB} population. Similar to the discussion in Section \ref{subsec:BHHMXB}, if non-conservative mass transfer leads to merging, it can help account for the observed absence of lower Be-star masses in \acp{BeXRB}, although this could be in tension with the relatively high observed \ac{BeXRB} formation rate in the Small Magellanic Cloud.

\subsubsection{Gravitational-wave sources}
\label{subsec:BBHs}
The formation of coalescing compact-object binaries, which are primary targets for current gravitational-wave detectors, is also sensitive to mass-transfer efficiency. At least 83 gravitational-wave events from \ac{BBH} coalescences have been reported by the LIGO-Virgo-KAGRA Collaboration upon completing their third observing run \citep{LVK+23_gwtc3}. A major channel for the isolated formation of \acp{BBH} involves stable case B or C mass transfer \cite[see, e.g.,][and references therein]{Mandel+Farmer22}.

We use \COMPAS (v02.41.04) to examine stellar and binary properties at the onset of case B or C mass transfer leading to a \ac{BBH} that merges within a Hubble time. We adopted \COMPAS's default model assumptions and a fixed metallicity of $Z=0.00142$, approximately one-tenth of the solar value. The typical case B donor is 32\Msun while the accretor is 25\Msun, separated by around 70\Rsun. Due to the short thermal timescale of the case B donor, we find extreme mass transfer rates ranging from a few times $10^{-2}$ up to $10^{-1}\Msunyr$. While the thermally-limited model results in $\beta \sim 0.1$, our prescription predicts extreme inflation and therefore nearly fully non-conservative mass transfer. This may lead to less massive secondary \acp{BH} being formed and greater orbital tightening. A self-consistent calculation is required to determine in detail the impact on the predicted detection rate and mass spectrum of \acp{BBH} mergers. Population synthesis calculations by \cite{Bouffanais+21} and \cite{vanSon+22} indicate that lowering the assumed mass-transfer efficiency decreases the predicted rates of merging \acp{BBH} and shifts the chirp mass spectrum towards lower masses.

\subsection{Observational counterparts}
\label{subsec:observations}
Semi-detached systems provide direct observational insights into binary mass transfer. An example is classical Algols, which contain less massive components (hereinafter, the secondary) that are more evolved. Thus, they are believed to be in a slower phase of nuclear timescale mass transfer following mass-ratio reversal. Conversely, some semi-detached binaries, also known as ``reverse Algols'', do not exhibit mass ratio inversion, possibly representing rarer systems undergoing shorter-lived, thermal timescale mass transfer before evolving into classical Algols \citep{Stepien11}. SV Cen and UX Mon are potential candidates for such systems \citep{Wilson+Starr76,Rucinski+92,Sudar+11}. Other systems, like $\beta$ Lyr and W Ser, could represent binaries observed shortly after mass-ratio reversal, where the mass-transfer rate is close to its maximum \citep{Deschamps+13}.

The accretor inflation we studied could give rise to near-contact binaries, which are semi-detached systems where the detached component exhibits a large filling fraction. This distinguishes them from classical Algols, which tend to have low filling factors for the primary. Near-contact binaries containing evolved primaries are classified as SD1 or V1010 Oph-type systems \citep{Yakut+Eggleton05}. Like reverse Algols, they potentially correspond to binaries in a rapid phase of mass transfer.

Higher-mass analogues of Algol-like binaries are rarer because mass transfer may terminate before mass-ratio reversal, eliminating a subsequent phase of nuclear-timescale mass transfer \citep{Sen+23}. The shorter thermal timescales of massive donors add to their scarcity. Nonetheless, recent works have modelled and searched for these systems. \cite{Sen+23} suggested VFTS 094 ($30.5+28\Msun$) and VFTS 176 ($28.3+17\Msun$) in the Tarantula Nebula as possible systems that experienced fast case-A mass transfer. Assuming conservative accretion, the initial donor-to-accretor thermal timescale ratio is expected to be of the order of $\sim10$ and $\sim100$ for VFTS 094 and VFTS 176, respectively \citep{Sen+23}. This could result in moderate accretor inflation that does not reach the Hayashi line. In cases with large thermal timescale ratios, the accretor could evolve into contact, providing a possible pathway towards forming contact binaries like W UMa variables and massive overcontact binaries.

\subsection{Possible effects of angular momentum accretion}
\label{subsec:rotation}
Although not modelled in our study, the angular momentum of accreted material is expected to significantly impact binary mass transfer. A \ac{MS} star can reach critical rotation after accreting a few percent of its initial mass. Indeed, accretors in longer-period Algols ($P_\mathrm{orb}\gtrsim 5~\mathrm{d}$) tend to rotate faster than their orbits \citep{Dervisoglu+10}. Mass transfer could also be responsible for the rapid rotation of Be stars in binaries.

The centrifugal support due to rotation will increase the size of the inflated accretor envelope. At the same time, the Roche potential needs to be adapted to account for supersynchronous rotation, which gives rise to a smaller ``rotational equipotential lobe'' \citep{Plavec58}. Consequently, accretor \ac{RLOF} is more likely, leading to non-conservative mass transfer in our framework. Notably, in so-called ``double contact binaries'', both stars are in contact with their Roche/critical surfaces, but at least one component rotates supersynchronously, meaning the stars are not physically touching. Proposed candidates for such systems include $\beta$ Lyr, V356 Sag, U Cep, and RZ Scu \citep{Wilson79}.

Observations have revealed the presence of accretion disks around the mass-gainers of $\beta$ Lyr, W Ser, UX Mon, and possibly SV Cen \citep{Linnell+Scheick91}. Bipolar outflows have also been observed in V356 Sag \citep{Peters+Polidan04} and originating from the supersynchronous component of $\beta$ Lyr \citep{Harmanec+96}. These structures remove angular momentum from the accretor, possibly enabling substantial mass gain if rotation is maintained below the critical rate \citep{Paczynski91,Popham+Narayan91}. Differential rotation in the inflated convective envelope could amplify existing magnetic fields, which could also spin down the accretor through magnetic braking and disk locking.

%% file: conclusion.tex
\section{Conclusions} \label{sec:conclusions}
We present fitting formulae for the maximum radii reached by ``hamstars'', stars accreting mass much faster than they may thermally accept. These conditions are encountered during cases B and C of mass transfer, where the evolved donor star has a much shorter thermal timescale than the accretor. Our fits are based on \MESA models of accreting \ac{ZAMS} stars, assuming spherically-symmetric and cold accretion at fixed rates.

Upon initiating accretion, the added material forms a tenuous, near-Eddington envelope that reaches hundreds to thousands of solar radii in extent. Beyond a critical accretion rate that depends on the initial stellar mass, an accretor expands along the Hayashi line and appears as a red giant. Contraction back towards the \ac{ZAMS} takes place once the accretor becomes luminous and large enough to transport away the excess gravitational energy of the envelope material. 

While this type of calculation had been performed more than four decades ago, the present study includes wider and denser sampling of initial stellar masses (2 to 20\Msun) and accretion rates (thirteen for each initial mass). This has enabled us to construct semi-empirical relations governing accretor evolution, including (i) an expression for the effective settling radius where gravitational energy of added material is released, and (ii) a power-law relation describing the luminosity along the Hayashi line as a function of mass. Along with the known temperature $\log(T~/~\mathrm{K}) \approx 3.6$ of the Hayashi limit, these relations allow the (maximum) luminosity and radius of an accretor on the Hayashi line to be estimated. We also provide fits for the maximum radii of models that contract before encountering the Hayashi limit.

We describe how to use these fits to estimate the mass-transfer efficiency given the mass donation rate and initial accretor mass, based on the assumption that fully conservative mass transfer may take place as long as the added mass fits inside the accretor Roche lobe. This prescription may be used in rapid binary population synthesis models, which can explore the impact of our results on predicted properties of stellar populations.

In Section \ref{subsec:examples}, we estimate the implications of our prescription for the formation of \ac{BH}-\acp{HMXB}, \acp{BeXRB}, and merging \acp{BBH}. Population synthesis studies have suggested that the observed masses and periods of \ac{BH}-\acp{HMXB} and \acp{BeXRB} require higher mass-transfer efficiencies than predicted by the conventional thermally-limited accretion model. However, the high mass transfer rates experienced by the progenitors of these binary systems result in considerable expansion, leading to highly non-conservative mass transfer within our framework. For merging \acp{BBH}, this may give rise to forming less massive secondary \acp{BH}. If extreme inflation leads to a merger, the scarcity of systems that have undergone inefficient accretion could contribute to better agreement with observations of \ac{BH}-\acp{HMXB} and \acp{BeXRB}.

We caution that our estimates can be sensitive to the conditions at the onset of \ac{RLOF}. A comprehensive population synthesis study should be performed to check the existence of parameter regions allowing for very efficient mass transfer. Our examples have centred on more massive stars. In contrast, low-mass accretors expand only mildly even when the accretion timescale is a factor of ten to a hundred times shorter than their thermal timescales. They may therefore experience more conservative accretion under our prescription, which holds potential significance for modelling systems such as Algol-type and hot subdwarf binaries. Mass transfer onto (partially) evolved accretors with shorter thermal timescales could also be conservative \citep[e.g.,][]{Maund:2004}. While simultaneously having an evolved donor and accretor requires nearly equal primary and secondary masses, this could be more likely following a previous episode of main-sequence (case A) mass transfer.

We also wish to highlight uncertainties and simplifications in our stellar models. They include the uncertain treatment of superadiabatic layers in near-Eddington envelopes, the addition of extra atmospheric pressure, and the assumption of cold, spherical accretion. Applying our results to non-spherical accretion requires a major assumption that accretion can be maintained near critical rotation, which was discussed in Section \ref{subsec:rotation}. It is therefore crucial to make progress in understanding the ability to regulate stellar rotation through an accretion disk, the role of tidal coupling between the inflated envelope and the orbit, and the impact of magnetic disk-locking. Notably, our calculations have also not considered the impact of a time-varying mass transfer rate or a more evolved accretor. As such, our study represents only an initial attempt to incorporate accretor expansion during binary mass transfer, but already constitutes an improvement upon existing prescriptions.